\documentstyle[preprint,aps]{revtex}

\begin{document}
\author{{\it E. J. Ferrer\thanks{%
e-mail: ferrer@fredonia.edu} and V. de la Incera\thanks{%
e-mail: incera@fredonia.edu}}}
\address{Dept. of Physics, State University of New York, Fredonia, NY 14063, USA
SUNY-FRE-96-04}
\title{SUPERCONDUCTIVITY IN ANYON FLUID AT FINITE TEMPERATURE AND DENSITY{\it
\thanks{%
Work presented at the DPF96 Meeting (Minneapolis, Aug.10-15, 1996). }}}
\maketitle

\begin{abstract}
The boundary effects in the screening of an applied magnetic field in a
charged anyon fluid at finite density ($\mu \neq 0$) and temperature ($T\neq
0$) are investigated. By analytically solving the extremum equations of the
system and minimizing the free energy density, we find that in a sample with
only one boundary (the half  plane), a total Meissner effect takes place;
while the sample with two boundaries (the infinite strip) exhibits a partial
Meissner effect. The short-range modes of propagation of the magnetic field
inside the fluid are characterized by two temperature dependent penetration
lengths.
\end{abstract}

\newpage

Since the claim by Laughlin and his collaborators\cite{1},\cite{2} that
fractional statistics could play a crucial role in high-T$_C$
superconductivity, a significant work has been done to investigate the
superconducting characteristics of the charged anyon fluid in two spatial
dimensions.

The anyon superconductivity at $T=0$ has been investigated by many authors%
\cite{2}-\cite{13a}. In this case, the anyon superconductivity appears due
to the exact cancellation between the bare and induced Chern-Simons terms in
the effective action of the theory.

The possible realization of anyon superconductivity at $T\neq 0$ has also
been extensively investigated\cite{8}-\cite{16}. At finite temperature,
based on non-vanishing correction to the induced Chern-Simons coefficient,
some authors (see, ref. \cite{9}) have concluded that the superconductivity
is lost at $T\neq 0$. In contrast with this result, in refs. \cite{11},\cite
{15} it was argued that the non-vanishing correction to the induced
Chern-Simons coefficient is numerically negligible at $T<200$ $^{\circ }K$.
On the other hand, the development of a pole $\sim \left( \frac{1}{{\bf k}%
^{2}}\right) $ at $T\neq 0$ in the polarization operator component $\Pi _{00}
$, characteristic of the Debye screening in plasmas, was found\cite{11},\cite
{15} as the main reason for the lack of a total Meissner effect in the
charged anyon fluid at finite temperature. In these papers it was discussed
how the appearance of this pole leads to a partial Meissner effect with a
penetration which appreciably increases with temperature. Independently, in
ref. \cite{8}, it was claimed that the anyon model fails to provide a good
superconducting behavior at finite temperature. The reason is that a
long-range mode was obtained inside the infinite bulk, which vanishes only
at $T=0$.

In the present paper, working in the self-consistent field approximation
\cite{8},\cite{11},\cite{15}, we show that the finite temperature
superconducting behavior of the charged anyon fluid depends on the sample
boundary conditions. This result is obtained by analytically solving the
field equations of the system and the stability conditions derived from the
free energy density, subject to two different set of boundary conditions: a
half infinite planar sample with an external magnetic field applied in the
boundary $\left( x=0\right) $, and an infinite strip with external magnetic
field applied in the two boundaries $\left( x=0\text{ and }x=L\right) $.

For the half plane we find that the external magnetic field cannot penetrate
the bulk (total Meissner screening). In this case the external magnetic
field is damped within the anyon fluid by two characteristic lengths,
corresponding to two short-range eigenmodes of propagation.

In the case of an infinite strip, it is shown that a partial penetration
occurs (partial Meissner screening). That is, the applied magnetic field
propagates, within the anyon fluid, through one long-range and two
short-range modes of propagation.

To understand the genesis of these results, one must take into account that
in this model the zero component of the Maxwell and Chern-Simons gauge
fields, $A_{0}$ and $a_{0}$ respectively, enter in the field equations in
the same foot as the electromagnetic and Chern-Simons field strengths.
Accordingly, $A_{0}$ and $a_{0}$ become physical, and, as we will prove,
their asymptotic behaviors (which are inherently linked to the sample
boundary conditions) affect the magnetic screening properties within the
bulk. In this sense, the model exhibits a kind of Aharonov-Bohm effect. The
importance of the boundary conditions in 2+1 dimensional models has been
already stressed in ref.\cite{b}.

The approach we follow is to compute the finite temperature effective action
starting from the Lagrangian density

\begin{equation}
{\cal L}=-\frac{1}{4}F_{\mu \nu }^{2}-\frac{N}{4\pi }\varepsilon ^{\mu \nu
\rho }a_{\mu }\partial _{\nu }a_{\rho }+en_{e}A_{0}+i\psi ^{\dagger
}D_{0}\psi -\frac{1}{2m}\left| D_{k}\psi \right| ^{2}+\psi ^{\dagger }\mu
\psi   \eqnum{1}  \label{1}
\end{equation}
of a 2+1 dimensional charged fluid of non-relativistic electrons, $\psi $,
coupled to two independent gauge fields, $A_{\mu }$ and $a_{\mu }$, which
represent the electromagnetic field and the Chern-Simons field respectively.
The covariant derivative is given by $D_{\nu }=\partial _{\nu }+i\left(
a_{\nu }+eA_{\nu }\right) ,\quad \nu =0,1,2.$ The charged character of the
fluid is implemented through the chemical potential $\mu $; $n_{e}$ is a
background neutralizing ``classical'' charge density. From the electric
charge neutrality condition, it is known that the system ground state has a
non-zero expectation value of the Chern-Simons magnetic field $\left(
\overline{b}=\frac{2\pi n_{e}}{N}\right) $.

To investigate the linear response of the medium to an applied external
magnetic field, it is enough to consider small fluctuations of the gauge
potentials around the many-particle ground state. That is, we can evaluate
the effective action corresponding to the Lagrangian density (1), up to
second order in these small quantities,

\begin{equation}
\Gamma _{eff}\,\left( A_\nu ,a_\nu \right) =\int dx\left( -\frac 14F_{\mu
\nu }^2-\frac N{4\pi }\varepsilon ^{\mu \nu \rho }a_\mu \partial _\nu a_\rho
+en_eA_0\right) +\Gamma ^{\left( 2\right) }  \eqnum{2}  \label{2}
\end{equation}

\begin{equation}
\Gamma ^{\left( 2\right) }=\int dx\Pi ^\nu \left( x\right) \left[ a_\nu
\left( x\right) +eA_\nu \left( x\right) \right] +\int dxdy\left[ a_\nu
\left( x\right) +eA_\nu \left( x\right) \right] \Pi ^{\mu \nu }\left(
x,y\right) \left[ a_\nu \left( y\right) +eA_\nu \left( y\right) \right]
\eqnum{3}  \label{3}
\end{equation}
$\Gamma ^{\left( 2\right) }$ is the fermion contribution to the effective
action in the above approximation, $\Pi _\nu $ and $\Pi _{\mu \nu }$
represent the fermion tadpole and polarization operators respectively. An
essential point in the study of this effective theory is the calculation of
these operators by using the fermion thermal Green's function defined in the
presence of the background field $\overline{b}$\cite{8},\cite{11}.

The leading behavior of these operators for static $\left( k_0=0\right) $
and slowly $\left( {\bf k}\sim 0\right) $ varying configurations have been
found in the low $\left( T\ll m/\overline{b}\right) $ and high $\left( T\gg
m/\overline{b}\right) $ temperature limits by different authors (see refs.
\cite{8},\cite{14}). These operators with the spatial momentum specialized
in the frame ${\bf k}=(k,0)$ are given by,

\begin{equation}
\Pi _{k}\left( x\right) =0,\qquad \qquad \Pi _{0}\left( x\right) =-n_{e}
\eqnum{4}  \label{4}
\end{equation}

\begin{equation}
\Pi _{\mu \nu }=\left(
\begin{array}{ccc}
{\it \Pi }_{{\it 0}}+{\it \Pi }_{{\it 0}}\,^{\prime }\,k^2 & 0 & {\it \Pi }_{%
{\it 1}}k \\
0 & 0 & 0 \\
-{\it \Pi }_{{\it 1}}k & 0 & {\it \Pi }_{\,{\it 2}}k^2
\end{array}
\right)  \eqnum{5}  \label{5}
\end{equation}
where the polarization operator coefficients in the different temperature
limits are,

{\it Low-Temperature Limit:}

\begin{equation}
{\it \Pi }_{{\it 0}}=\frac{\beta \overline{b}}\pi e^{-\beta \overline{b}/2m}
\eqnum{6}  \label{6}
\end{equation}

\begin{equation}
{\it \Pi }_{{\it 0}}\,^{\prime }=\frac{mN}{2\pi \overline{b}}\left[ 1-\frac{%
2\beta \overline{b}}me^{-\beta \overline{b}/2m}\right]  \eqnum{7}  \label{7}
\end{equation}

\begin{equation}
{\it \Pi }_{{\it 1}}=\frac N{2\pi }\left[ 1-\frac{2\beta \overline{b}}m%
e^{-\beta \overline{b}/2m}\right]  \eqnum{8}  \label{8}
\end{equation}

\begin{equation}
{\it \Pi }_{\,{\it 2}}=\frac{N^2}{2\pi m}\left[ 1+\frac 2{N^2}e^{-\beta
\overline{b}/2m}-\frac{2\beta \overline{b}}m\left( 1+\frac 1{4N^2}\right)
e^{-\beta \overline{b}/2m}\right]  \eqnum{9}  \label{9}
\end{equation}

{\it High-Temperature Limit:}

\begin{equation}
{\it \Pi }_{{\it 0}}=-\frac m{4\pi }\left[ \tanh \left( \frac{\beta \mu }2%
\right) +1\right]  \eqnum{10}  \label{10}
\end{equation}

\begin{equation}
{\it \Pi }_{{\it 0}}\,^{\prime }=-\frac \beta {96\pi }%
\mathop{\rm sech}%
{}^2\left( \frac{\beta \mu }2\right)  \eqnum{11}  \label{11}
\end{equation}

\begin{equation}
{\it \Pi }_{{\it 1}}=\frac{i\beta \overline{b}}{96\pi m}%
\mathop{\rm sech}%
{}^2\left( \frac{\beta \mu }2\right)  \eqnum{12}  \label{12}
\end{equation}

\begin{equation}
{\it \Pi }_{\,{\it 2}}=\frac 1{48\pi m}\left[ \tanh \left( \frac{\beta \mu }2%
\right) +1\right]  \eqnum{13}  \label{13}
\end{equation}

Another important quantity needed to investigate the behavior of the anyon
fluid is its free-energy

\[
{\cal F}=\frac 12\int\limits_{-L_2/2}^{L_2/2}dy\int\limits_0^{L_1}dx\left\{
\left( E^2+B^2\right) +{\it \Pi }_{{\it 0}}\left( eA_0+a_0\right) ^2+{\it %
\Pi }_{{\it 0}}\,^{\prime }\left( eA_0+a_0\right) \partial _x^2\left(
eA_0+a_0\right) \right.
\]

\begin{equation}
\left. +{\it \Pi }_{{\it 1}}\left[ \left( eA_0+a_0\right) \partial _x\left(
eA_2+a_2\right) -\left( eA_2+a_2\right) \partial _x\left( eA_0+a_0\right)
\right] +{\it \Pi }_{\,{\it 2}}\left( eA_2+a_2\right) \partial _x^2\left(
eA_2+a_2\right) \right\}  \eqnum{14}  \label{14}
\end{equation}

To study the linear response of the anyon fluid to an applied external
magnetic field we have to solve the extremum equations derived from the
effective action (2). This formulation is known in the literature as the
self-consistent field approximation\cite{11},\cite{15}. The corresponding
Maxwell and Chern-Simons extremum equations are respectively,

\begin{equation}
\partial _\nu F^{\nu \mu }=eJ_{ind}^\mu  \eqnum{15}  \label{15}
\end{equation}

\begin{equation}
-\frac N{4\pi }\varepsilon ^{\mu \nu \rho }f_{\nu \rho }=J_{ind}^\mu
\eqnum{16}  \label{16}
\end{equation}
Here, $f_{\mu \nu }$ is the Chern-Simons gauge field strength tensor,
defined as $f_{\mu \nu }=\partial _\mu a_\nu -\partial _\nu a_\mu $, and $%
J_{ind}^\mu $ is the current density induced by the anyon system at finite
temperature and density. Their different components are given by

\begin{equation}
J_{ind}^0\left( x\right) ={\it \Pi }_{{\it 0}}\left[ a_0\left( x\right)
+eA_0\left( x\right) \right] +{\it \Pi }_{{\it 0}}\,^{\prime }\partial
_x\left( {\cal E}+eE\right) +i{\it \Pi }_{{\it 1}}\left( b+eB\right)
\eqnum{17}  \label{17}
\end{equation}

\begin{equation}
J_{ind}^{1}\left( x\right) =0,\qquad J_{ind}^{2}\left( x\right) =i{\it \Pi }%
_{{\it 1}}\left( {\cal E}+eE\right) +{\it \Pi }_{\,{\it 2}}\partial
_{x}\left( b+eB\right)  \eqnum{18}  \label{18}
\end{equation}
In the above expressions the notation: ${\cal E}=f_{01}$, $E=F_{01}$, $%
b=f_{21}$ , $B=F_{21}$was used. Eqs. (17)-(18) play the role in the anyon
fluid of the London equations in BCS superconductivity. When the induced
currents (17), (18) are substituted in eqs. (15) and (16), one finds, after
some manipulation, the following set of independent differential equations,

\begin{equation}
\omega \partial _x^2B+\alpha B=\gamma \left[ \partial _xE-\sigma A_0\right]
+\tau \,a_0  \eqnum{19}  \label{19}
\end{equation}

\begin{equation}
\partial _xB=\kappa \partial _x^2E+\eta E  \eqnum{20}  \label{20}
\end{equation}

\begin{equation}
\partial _xa_0=\chi \partial _xB  \eqnum{21}  \label{21}
\end{equation}
The coefficients appearing in these differential equations depend on the
components of the polarization operators through the relations,

\[
\omega =\frac{2\pi }N{\it \Pi }_{{\it 0}}\,^{\prime },\quad \alpha =ie^2{\it %
\Pi }_{{\it 1}},\quad \tau =-e{\it \Pi }_{{\it 0}},\quad \chi =-\frac{2\pi }{%
eN},\quad \sigma =\frac{e^2}\gamma {\it \Pi }_{{\it 0}},\quad \eta =-\frac{%
ie^2}\delta {\it \Pi }_{{\it 1}}
\]

\begin{equation}
\gamma =1-e^{2}{\it \Pi }_{{\it 0}}\,^{\prime }-\frac{2\pi i}{N}{\it \Pi }_{%
{\it 1}},\quad \delta =1+e^{2}{\it \Pi }_{\,{\it 2}}+\frac{2\pi i}{N}{\it %
\Pi }_{{\it 1}},\quad \kappa =-\frac{2\pi }{N\delta }{\it \Pi }_{\,{\it 2}}.
\eqnum{22}  \label{22}
\end{equation}

The extremum equations (19)-(21) are not essentially different from those
found for the anyon effective theory at finite temperature by other authors%
\cite{11},\cite{15}. Distinctive of these equations is the appearance of the
nonzero constant coefficients $\sigma $ and $\tau $. They are related to the
Debye screening which is a property of the charged medium. It is a peculiar
fact that in the anyon fluid these coefficients appear linked to the
magnetic field (see eq. (19)). As a consequence, the zero components of the
gauge potentials, $A_{0}$ and $a_{0}$, play a nontrivial role in the
magnetic field solution of eqs. (19)-(21).

To solve eqs. (19)-(21) we can conveniently arrange them to obtain,

\begin{equation}
a\partial _x^4E+d\partial _x^2E+cE=0  \eqnum{23}  \label{23}
\end{equation}
where $a=\omega \kappa $, $d=\omega \eta +\alpha \kappa -\gamma -\tau \kappa
\chi $, and $c=\alpha \eta -\sigma \gamma -\tau \eta \chi $. Then the
solutions for the fields $E$, and $B$, and for the potentials $a_0$ and $A_0$%
, can be obtained from (23), (20), (21) and the definition of $E$ in terms
of $A_0$, respectively. Being (23) a higher order differential equation, its
solution belongs to a wider class if compared to that corresponding to the
original eqs. (19)-(21). Thus, to exclude the redundant solutions we have to
require that they satisfy eq. (19) as a supplementary condition. In this way
we can reduce the number of independent unknown coefficients to six, which
is the number corresponding to the original system (19)-(21).

Solving eq. (23) we obtain,

\begin{equation}
E\left( x\right) =C_1e^{-x\xi _1}+C_2e^{x\xi _1}+C_3e^{-x\xi _2}+C_4e^{x\xi
_2},  \eqnum{24}  \label{24}
\end{equation}
where

\begin{equation}
\xi _{1,2}=\left[ -d\pm \sqrt{d^2-4ac}\right] ^{\frac 12}/\sqrt{2a}
\eqnum{25}  \label{25}
\end{equation}
take real values at any temperature when evaluated with the typical values $%
n_e=\left( 1\sim 5\right) \times 10^{14}cm^{-2}$, $m=2m_e$ ($m_e=2.6\times
10^{10}cm^{-1}$ is the electron mass) and $\left| N\right| =2$.

With the solution (24), eqs. (20), (21), and the definition $E=-\partial
_xA_0$, we find,

\begin{equation}
B\left( x\right) =\gamma _1\left( C_1e^{-x\xi _1}-C_2e^{x\xi _1}\right)
+\gamma _2\left( C_3e^{-x\xi _2}-C_4e^{x\xi _2}\right) +C_5  \eqnum{26}
\label{26}
\end{equation}

\begin{equation}
a_0\left( x\right) =\chi \gamma _1\left( C_1e^{-x\xi _1}-C_2e^{x\xi
_1}\right) +\chi \gamma _2\left( C_3e^{-x\xi _2}-C_4e^{x\xi _2}\right) +C_6
\eqnum{27}  \label{27}
\end{equation}

\begin{equation}
A_0\left( x\right) =\frac 1{\xi _1}\left( -C_1e^{-x\xi _1}+C_2e^{x\xi
_1}\right) +\frac 1{\xi _2}\left( -C_3e^{-x\xi _2}+C_4e^{x\xi _2}\right) +C_7
\eqnum{28}  \label{28}
\end{equation}
Corresponding to the magnetic field (26) we have the electromagnetic
potential $A_2$ given by,

\begin{equation}
A_2\left( x\right) =-\frac{\gamma _1}{\xi _1}\left( C_1e^{-x\xi
_1}-C_2e^{x\xi _1}\right) -\frac{\gamma _2}{\xi _2}\left( C_3e^{-x\xi
_2}-C_4e^{x\xi _2}\right) +C_5x  \eqnum{29}  \label{29}
\end{equation}
The spatial component of the Chern-Simons field is

\begin{equation}
a_2\left( x\right) =\chi \left( C_1e^{-x\xi _1}+C_2e^{x\xi _1}+C_3e^{-x\xi
_2}+C_4e^{x\xi _2}\right)  \eqnum{30}  \label{30}
\end{equation}
In the above formulas we introduced the notation,

\begin{equation}
\gamma _1=\frac{\xi _1^2\kappa +\eta }{\xi _1},\qquad \quad \gamma _2=\frac{%
\xi _2^2\kappa +\eta }{\xi _2}  \eqnum{31}  \label{31}
\end{equation}

The extra unknown coefficient is eliminated, as it was explained above,
substituting the solutions (24), (26), (27) and (28) into eq. (19) to obtain
the relation,

\begin{equation}
C_5=\frac \tau \alpha C_6+\frac{\sigma \gamma }\alpha C_7  \eqnum{32}
\label{32}
\end{equation}
The last relation establishes a connection between the asymptotic conditions
for the zero components of the gauge potentials and the asymptotic condition
for the magnetic field.

Let us take into account now the boundary conditions needed to determine the
six independent unknown coefficients. Henceforth, we consider two different
sample configurations: the half plane and the infinite strip.

{\it The half plane:}

We will consider the anyon fluid confined to a semi-infinite plane $-\infty
<y<\infty $ with boundary at $x=0$. The external magnetic field will be
applied from the vacuum ($-\infty <x<0$). We restrict our solution to gauge
field configurations which are static and uniform in the $y$-direction.

The boundary conditions for the magnetic field are $B\left( x=0\right) =%
\overline{B}$ ($\overline{B}$ constant), and $B\left( x\rightarrow \infty
\right) $ $finite$. Because no external electric field is applied, the
boundary conditions for this field are, $E\left( x=0\right) =0$, $E\left(
x\rightarrow \infty \right) $ $finite$.

With the above conditions we obtain $C_2=C_4=0$ and $C_1=-C_3$, where $C_1$
depends on the magnetic field boundary value, $\overline{B}$, the unknown
coefficient $C_5$ and temperature, through the relation,

\begin{equation}
C_1=\frac{\overline{B}-C_5}{\gamma _1-\gamma _2}  \eqnum{33}  \label{33}
\end{equation}

To find the remaining independent unknown coefficients ($C_6$ and $C_7$) we
will consider the system stability condition. That is, starting from the
system free energy (14) evaluated in the field solutions (24), (26)-(30), we
define the free energy density ${\it f}=\frac{{\cal F}}{{\cal A}}$, where
the area of the sample is given by ${\cal A}=L_1L_2$. Then, we find the
values of $C_6$ and $C_7$ that minimize the free-energy density,
\begin{equation}
\frac{\delta {\it f}}{\delta C_6}=0,\text{ \qquad \qquad }\frac{\delta {\it f%
}}{\delta C_7}=0  \eqnum{34}  \label{34}
\end{equation}

Considering the leading terms appearing in eq. (34) after taking the half
plane limit ($L_{1}\rightarrow \infty $, $L_{2}\rightarrow \infty $), we
arrive to the following equations,

\begin{equation}
C_5+\frac 12{\it \Pi }_{{\it 1}}\left( eC_7+C_6\right) =0  \eqnum{35}
\label{35}
\end{equation}

\begin{equation}
C_5+\frac{2{\it \Pi }_{{\it 0}}}{{\it \Pi }_{{\it 1}}}\left( eC_7+C_6\right)
=0  \eqnum{36}  \label{36}
\end{equation}
From eqs. (35) and (36), together with the constraint (32), we obtain $%
C_5=C_6=C_7=0$.

The magnetic field penetration is then given by,

\begin{equation}
B\left( x\right) =B_{1}\left( T\right) \,e^{-x\xi _{1}}-B_{2}\left( T\right)
\,e^{-x\xi _{2}},\;\hspace{0.3in}x\geq 0  \eqnum{37}  \label{37}
\end{equation}
where the temperature dependent coefficients, $B_{1}\left( T\right) $ and $%
B_{2}\left( T\right) $, are given by,

\begin{equation}
B_{1}\left( T\right) =\frac{\gamma _{1}}{\gamma _{1}-\gamma _{2}}\overline{B}%
,\qquad B_{2}\left( T\right) =\frac{\gamma _{2}}{\gamma _{1}-\gamma _{2}}%
\overline{B}  \eqnum{38}  \label{38}
\end{equation}

Hence, the applied magnetic field within the anyon fluid totally falls down
exponentially on two essentially different scales, $\lambda _{1}=1/\xi _{1}$
and $\lambda _{2}=1/\xi _{2}$, which characterize two eigenmodes of
propagation inside the fluid. Considering the obtained values for the $C_{i}$
coefficients in the solution (24), we also find that the induction of an
electric field inside the bulk is intimately linked to the Meissner effect
in the anyon fluid. Note that this induced electric field also decays
exponentially within the characteristic lengths $\lambda _{1},\lambda _{2}.$

{\it The infinite strip:}

Now we consider the anyon fluid confined to the strip $-\infty <y<\infty $
with boundaries at $x=0$ and $x=L_{1}=2L$. The external magnetic field will
be applied from the vacuum ($-\infty <x<0$, $2L<x<\infty $). We again
restrict our solution to gauge field configurations which are static and
uniform in the $y$-direction.

We consider the following symmetric boundary conditions for the magnetic
field: $B\left( x=0\right) =\overline{B}$, $B\left( x=2L\right) =\overline{B}
$ ($\overline{B}$ constant). The boundary conditions for the electric field
are, $E\left( x=0\right) =0$, $E\left( x=2L\right) =0$. With the above
boundary conditions and considering that $L\gg \lambda _1$, $\lambda _2$, we
obtain that the unknown coefficients $C_1$, $C_2$, $C_3$, $C_4$ and $C_5$,
are related through the following equations,

\[
C_1=\frac{C_5-\overline{B}}{\gamma _1-\gamma _2}
\]

\begin{equation}
C_2=-C_1e^{-2L\xi _1},\qquad \quad C_3=-C_1,\quad \qquad C_4=C_1e^{-2L\xi _2}
\eqnum{39}  \label{39}
\end{equation}

To determine the independent unknown coefficients, $C_{6}$ and $C_{7}$, we
repeat the same procedure we used in the half plane case. Taking $L_{1}=2L$
in the free energy (14), we find that in the $L\gg \lambda _{1}$, $\lambda
_{2}$ limit, the leading terms appearing in eqs. (34), are

\begin{equation}
C_6={\cal K}_2\overline{B},\qquad \quad C_7={\cal K}_1C_6  \eqnum{40}
\label{40}
\end{equation}
where

\begin{equation}
{\cal K}_1=\left( {\it \Pi }_{{\it 0}}+\frac{\kappa \gamma }{2\alpha }{\it %
\Pi }_{{\it 1}}\right) ^{-1}\left( {\it \Pi }_{{\it 0}}+\frac{e\tau }{%
2\alpha }{\it \Pi }_{{\it 1}}\right)  \eqnum{41}  \label{41}
\end{equation}

\begin{equation}
{\cal K}_2=\left[ \left( \frac{\gamma _1-\gamma _2}{\widetilde{\Pi }}%
+2\right) \left( \frac{\kappa \gamma }{\alpha {\cal K}_1}+\frac \tau \alpha
\right) +\frac{e{\it \Pi }_{{\it 1}}\left( \gamma _1-\gamma _2\right) }{2%
\widetilde{\Pi }}\left( \frac e{{\cal K}_1}+1\right) \right] ^{-1}
\eqnum{42}  \label{42}
\end{equation}

\begin{equation}
\widetilde{\Pi }=\left[ e\chi \left( \gamma _2-\gamma _1\right) +e^2\left(
\frac 1{\xi _2}-\frac 1{\xi _1}\right) \right] \frac{{\it \Pi }_{{\it 1}}}2%
+\left[ e^2\left( \gamma _2-\gamma _1\right) +\frac{2\pi }N\left( \xi _2-\xi
_1\right) \right] \frac{{\it \Pi }_{{\it 2}}}2  \eqnum{43}  \label{43}
\end{equation}

From (40) and the constraint equation (32), we have that $C_5$ is not zero,
what implies a partial Meissner effect. In this case the magnetic field
inside the anyon fluid is given by

\begin{equation}
B\left( x\right) =\overline{B}_1\left( T\right) \left( e^{-x\xi
_1}+e^{-\left( 2L-x\right) \xi _1}\right) -\overline{B}_2\left( T\right)
\left( e^{-x\xi _2}+e^{-\left( 2L-x\right) \xi _2}\right) +\overline{B}%
\left( T\right)  \eqnum{44}  \label{44}
\end{equation}
where

\begin{equation}
\overline{B}_1\left( T\right) =\frac{\gamma _1}{\gamma _1-\gamma _2}\left(
\overline{B}-\overline{B}\left( T\right) \right) ,\qquad \overline{B}%
_2\left( T\right) =\frac{\gamma _2}{\gamma _1}\overline{B}_1\left( T\right)
,\qquad \overline{B}\left( T\right) ={\cal K}_2\left( \frac{\kappa \gamma }%
\alpha {\cal K}_1+\frac \tau \alpha \right) \overline{B}  \eqnum{45}
\label{45}
\end{equation}

At $x=L$ (i.e., in the middle of the sample), considering the large $L$
limit, we have that $B\left( x=L\right) \simeq \overline{B}\left( T\right) $%
. This can be interpreted as a partial Meissner effect, taking into account
that $\overline{B}\left( T\right) \leq \overline{B}$.

The results of this paper explicitly show something we had previously
pointed out \cite{16}, namely, that in the charged anyon fluid the zero
components of the gauge potentials become physical. Their asymptotic
behavior, which through the equations of motion affect the magnetic field in
the bulk, are fixed by the conditions of minimal free energy density and by
the sample boundary conditions. The physical relevance of gauge potentials
is not new in Field Theory. Indeed, in statistical gauge theory it is
natural to expect that different asymptotic behaviors of the zero components
of the gauge fields correspond to different physical situations, since it is
known that non-zero constant asymptotic gauge field configurations are not
gauge equivalent (under proper, periodic gauge transformations) to the
trivial vacuum\cite{20}. The system under study here is just an example of
such a case.

Acknowledgments: The authors are very grateful for stimulating discussions
with Profs. G. Baskaran, A. Cabo, E.S. Fradkin, Y. Hosotani and J.
Strathdee. We would also like to thank Prof. S. Randjbar-Daemi for kindly
bringing the publication of ref. \cite{b} to our attention. Finally, it is a
pleasure for us to thank Prof. Yu Lu for his hospitality during our stay at
the ICTP, where part of this work was done. This research has been supported
in part by the National Science Foundation under Grant No. PHY-9414509.


\begin{references}
\bibitem{1}  V. Kalmeyer and R. B. Laughlin, Phys. Rev. Lett. {\bf 59}, 2095
(1987); R. B. Laughlin, Phys. Rev. Lett. {\bf 60}, 2677 (1988).

\bibitem{2}  A. L. Fetter, C. B. Hanna and R. B. Laughlin, Phys. Rev. {\bf %
B39}, 9679 (1989).

\bibitem{3}  C.B. Hanna, R.B. Laughlin and A.L. Fetter, Phys. Rev. {\bf B40}%
, 8745 (1989), {\it ibid.}{\bf \ 43}, 309 (1991); G.S. Canright, S.M. Girvin
and A. Brass, Phys. Rev. Lett. {\bf 63}, 2291, 2295 (1989); X.G. Wen and A.
Zee, Phys. Rev. {\bf B41}, 240 (1990); E. Fradkin, Phys. Rev. Lett. {\bf 63}%
, 322 (1989); Phys. Rev. {\bf B42}, 570 (1990); Y. Hosotani and S.
Chakravarty, Phys. Rev. {\bf B42}, 342 (1990); Phys. Rev. {\bf D44}, 441
(1991); A. L. Fetter and C. B. Hanna, Phys. Rev. {\bf B45}, 2335 (1992).

\bibitem{4}  Y.-H Chen, F. Wilczek, E. Witten and B. Halpering,Int . J. Mod.
Phys. {\bf B3}, 1001 (1989).

\bibitem{8}  S. Randjbar-Daemi, A. Salam and J. Strathdee, Nucl. Phys. {\bf %
B340}, 403 (1990).

\bibitem{9}  J. D. Lykken, J. Sonnenschein and N. Weiss, Phys. Rev. {\bf D42}%
, 2161 (1990); Int. J. Mod. Phys. {\bf A6}, 1335 (1991).

\bibitem{11}  Y. Hosotani, Int. J. Mod. Phys. {\bf B7}, 2219 (1993).

\bibitem{13a}  E. J. Ferrer and V. de la Incera, Int. J. Mod. Phys. {\bf B9}%
, 3585 (1995).

\bibitem{13}  P. K. Panigrahi, R. Ray and B Sakita, Phys. Rev. {\bf B42},
4036 (1990); J. Kapusta, M. E. Carrington, B. Bayman, D. Seibert and C.S.
Song, Phys. Rev. {\bf B44}, 7519 (1991); Y. Georgelin, M. Knecht, Y.
Leblanc, and J. C. Wallet, Mod. Phys. Lett. {\bf B5}, 211 (1991); Y.
Leblanc, and J. C. Wallet, Mod. Phys. Lett. {\bf B6}, 1623 (1992); I. E.
Aronov, E.N. Bogachek, I. V. Krive and S. A. Naftulin, JETP Lett. {\bf 56},
283 (1992); Y. Kitazawa and H. Murayama, Phys. Rev. {\bf B41}, 11101 (1990).

\bibitem{14}  S. S. Mandal, S. Ramaswamy and V. Ravishankar; Mod. Phys.
Lett. {\bf B8}, 561 (1994), Int. J. Mod. Phys. {\bf B8}, 3095 (1994).

\bibitem{15}  J. E. Hetric, Y. Hosotani and B.-H Lee, Ann. Phys {\bf 209},
151 (1991); J. E. Hetric and Y. Hosotani, Phys. Rev. {\bf B45}, 2981 (1992).

\bibitem{16}  E. J. Ferrer, R. Hurka and V. de la Incera, {\it %
``High-Temperature Anyon Superconductivity'', Fredonia preprint
SUNY-FRE-96-02,}{\bf \ hep-th/9602042}.

\bibitem{b}  S. Randjbar-Daemi, A. Salam and J. Strathdee, Int. J. Mod.
Phys. {\bf B5} (1991) 845.

\bibitem{20}  N. Batakis and G. Lazarides, Phys. Rev. D {\bf 18}, 4710
(1978);D.J. Gross, R. D. Pisarski, and L. G. Yaffe, Rev. Mod. Phys. {\bf 53}%
, 43 (1981); A. Actor, Phys. Rev. D {\bf 27}, 2548 (1983); Ann. Phys {\bf 159%
}, 445 (1985).
\end{references}
\end{document}